# Enhancing the Usability of Self-service Kiosks for Older Adults: Effects of Using Privacy Partitions and Chairs

## Hyesun Chung

Department of Industrial Engineering, Seoul National University, Seoul, South Korea

## Woojin Park

Department of Industrial Engineering and Institute for Industrial Systems Innovation, Seoul National University, Seoul, South Korea

**Contact information for corresponding author**

Woojin Park

Professor

Department of Industrial Engineering, Seoul National University

1 Gwanak-ro, Gwanak-gu, Seoul 08826, South Korea

woojinpark@snu.ac.kr

82-2-880-4310 (Office)

82-10-2075-9468 (Cell phone)

82-2-889-8560 (Fax)

# Abstract

**Objective:** This study aimed to evaluate the effects of possible physical design features of self-service kiosks (SSK), privacy partitions and chairs, on workload and task performance of older users during a typical SSK task.

**Background:** Actionable design guidelines for lessening physical demands, raising confidence toward SSK, and eliminating situational deterrents for older users are lacking.

**Method:** The current study comparatively evaluated eight physical SSK design alternatives, which were the combinations of two levels (Yes/No) of three physical design variables: Side Partition, Back Partition, Chair. Younger (n = 22) and older (n = 27) participants performed a menu ordering task using each physical design alternative.

**Results:** Older participants showed a large variation in task performance across the design alternatives indicating stronger impacts of the physical design features. In particular, task completion time was significantly shorter while the average reaction time was longer when they were seated than standing. Sitting also reduced workload in multiple dimensions, including time pressure and frustration. In addition, the use of either side or back partitions reduced mean ratings of mental demand and effort.

**Conclusion:** Physical design features that can lower divided attention costs or assist postural control would benefit older adults during the use of SSK. The study suggests placing chairs and either side or back partitions to enhance older adults's user experience.

**Application:** The physical design guidelines derived in this study are expected to enhance the usability of SSK for older adults and thereby contribute to increasing their social engagement and quality of life.

## Précis

A laboratory experiment was conducted to evaluate the effects of using privacy partitions and chairs on workload and task performance when using self-service kiosks (SSK). Two age groups, younger and older, were compared in the effects of the physical design features. Making use of chairs and either side or back partitions was beneficial, especially for older adults.
**3 / 43**

# INTRODUCTION

Since the advent of automated teller machines (ATMs), self-service technologies (SST) have been increasingly adopted. The main advantages they provide include time and cost savings, greater control over service delivery, and convenience (Meuter & Bitner, 1998; Kauffman & Lally, 1994; Dabholkar, 1996; Bitner et al., 2000; Salomann et al., 2006). Currently, they are widely available across diverse service organizations, such as fast-food restaurants, banks, hospitals, information centers, and more (Hagen & Sandnes, 2010); and, a significant portion of the population has some form of SST use experience (Wang et al., 2012), including the interaction with self-service kiosks (SSK). The deployment of SST would increase further as consumers continue to embrace SST, and businesses rush to reap the benefits (Kiosk Marketplace, 2019). The interactive kiosk market is expected to grow from USD 26.2 billion in 2020 to USD 32.8 billion by 2025, at a CAGR of 4.6% (Markets and Markets, 2020).

Despite their benefits and utilities, however, SST also came with their own problems. One such problem is the difficulties experienced by older adults during the use of SSK (e.g., Park, 2019; Miller, 2019; Poulter, 2017). Many older adults users find it difficult to read text information displayed on SSK (Hagen & Sandnes, 2010), figure out the correct user action sequence for a given SSK task (Barnard et al., 2013), and perform prolonged standing while conducting an SSK task (Park, 2019). Many describe the experience as stressful, unpleasant, and physically demanding. These difficulties can be largely attributed to the failure in considering the personal, attitudinal and situational characteristics of older adults users in the design of SSK. Aging is associated with significant declines in perceptual, cognitive and motor control abilities, and muscular strengths (Iancu, I. & Iancu, B., 2017; Caparni et al., 2012). Older adults are also characterized by technology anxiety and lack of confidence towards technologies that negatively affect task performance during the use of technologies (e.g., Dean, 2008; Wang et al., 2012; Gelbrich & Sattler, 2014). These attitudes are known to be invoked by situational stressors, such as time pressure and presence of others.

The difficulties described above represent a significant problem as they can seriously compromise the functioning and well-being of older adults in everyday life. Indeed, they should be understood as barriers that exacerbate the current generational digital divide for older adults population (Van Dijk, 2006). The digital divide, defined as inequities in terms of who accesses and benefits from the digital landscape (Fang et al., 2019), is deeply related to decreased social engagement and lower life independence (Hill et al., 2015). It is also known to produce a new form of social isolation referred to as "digital exclusion" (Seifert et al., 2018). Hill et al. (2015) described digital exclusion as a "cumulative, self-propelling spiral of isolation whereby the digitally rich continue to become included and the digitally poor continue to become isolated within a culture where more of society's business and culture is conducted through technology."

Thus, in human factors and related fields, much research has attempted to improve the usability of SSK by creating designs that accommodate the characteristics of older adults users. The suggested solutions mainly involved designing clear inputs and outputs with the use of proper graphical user interface elements (e.g., color, text size, button size and spacing) and feedback, and/or adjusting the parameters of the touchscreen interfaces (e.g., display color, contrast ratio, inclination angle, height) according to the user's body dimensions (e.g., Charness et al., 2010; Hawthorn, 2000; Caparni et al., 2012; Tsai & Lee, 2009; Hwangbo et al., 2012; Colle & Hiszem, 2004; Gao & Sun, 2015; Sesto et al., 2012; Kim et al., 2014; Jin et al., 2007; Murata & Iwase, 2005; Schedlbauer, 2007; Hagen & Sandnes, 2010).

Despite the past research efforts, however, many of the difficulties that older adults users experience during the use of SSK remain unaddressed at this time. Design solutions or guidelines for lessening muscular loadings, raising confidence, and eliminating situational stresses for older adults do not seem available. Also, little research has been conducted concerning these issues.

One promising approach to address the unresolved difficulties above is innovating the physical design of SSK. The term "physical design" here refers to deciding upon the physical form factor design of SSK. It involves specifying physical design parameters and also making decisions on the inclusion of physical design features, such as privacy partitions and a chair.

The physical design is thought to represent an opportunity for enhancing the user experience through simple design improvements. For example, adding side partitions and a chair to SSK may help the older users with balance control as the two physical design features would work as visual references and physical support, respectively. It is possible to hypothesize that such reductions in balance control demands would free up cognitive resources for older adults users so that they can better concentrate on the main SSK task – multiple dual-task studies involving older adults participants (Lacour et al., 2008; Marsh & Geel, 2000; Verhaeghen et al., 2003) point to this possibility. Also, installing side and back partitions would likely mitigate the situational deterrents such as time pressure and may help alleviate the older users' anxiety and boost their confidence toward using SSK. Making use of a chair may also help reduce time pressure, thus leading them to a more confident state. All of these changes may lead to better performance because a high level of self-efficacy would activate the cognitive and intellectual functioning, and consequently improve task performance (Bandura et al., 1999).

Thus, in an effort to contribute to improving the experience of the older users during the use of SSK and further eliminating the digital divide, the objective of the current study was to explore possible benefits of using three physical design features (side partitions, a back partition, and a chair) that can be easily added to SSK. In particular, the study aimed to determine if the physical design features would reduce workload and improve task performance of older users during a typical SSK task, and investigate if they benefit younger users in similar ways. The study also attempted to evaluate eight physical design alternatives in terms of task performance and workload. On the basis of the experimental results, this study provided design guidelines on the utilization of the physical design features.

# LITERATURE REVIEW

## Characteristics of the aged population

The difficulties that older adults users experience during the use of SSK can largely be attributed to their cognitive, physical, attitudinal and situational characteristics. Some of the relevant characteristics and their impacts are described in what follows:

**Cognitive characteristics**

Older adults experience declines in visual acuity (Gao & Sun, 2015). The reduced vision is the major challenge as most user interface elements of SSK are visual (Caparni et al., 2012).

Age-associated changes in the cognitive system are also characterized by the reduction of cognitive resources available (Cerella, 1985), and, in particular, the declines in memory, executive functions, and speed of information processing (Deary et al., 2009). Many technological systems, including SSK, rely on a person's ability to keep information active; however, this could be demanding for many older users (Caparni et al., 2012).

Older adults also typically do not have prior experiences of or mental models for using SSK or similar systems - learning to control a system requires constructing an adequate mental model that describes the causal connections between user actions, system responses, and user goals (Lewis, 1986). Acquiring new mental models for using systems with high degrees of freedom, such as SSK, is difficult for inexperienced older adults (Barnard et al., 2013).

**Physical characteristics**

Older adults are generally smaller in stature relative to younger adults (Czaja et al., 2019). SSK designed without considering the anthropometric characteristics can disaccommodate a large portion of older adults population.

Older adults also experience declines in motor control, dexterity, speed of execution, hand-eye coordination, mobility, and agility (Iancu, I. & Iancu, B., 2017). Aging also has a degenerative effect on hand functions, including declines in manual speed and finger dexterity (Ranganathan et al., 2001), thus hindering them from making precise selections of small interface targets (Caparni et al., 2012).

Balance and coordination capabilities (Teasdale et al., 1993) as well as muscular strengths (Caparni et al., 2012) decrease as a function of age. Consequently, it is harder for older adults to perform physically demanding tasks, including prolonged standing.

**Attitudinal and situational characteristics**

Older adults are generally less confident in utilizing technology, and their under-confidence is a significant source of difficulties that they encounter in mastering new technology (Marqui et al., 2002). In fact, according to the survey by Dean (2008), older consumers had fewer experiences of SST and less confidence in using them. They tended to think that they might make mistakes that could not be corrected by themselves (Wang et al., 2012). As self-efficacy positively correlates with the intention to use technology (Walker & Johnson, 2006; Lee & Coughlin, 2015; Peral-Peral et al., 2019) and SST (Simon & Usunier, 2007; Blut et al., 2016; Wang, 2017), increasing older adults's confidence and perceived self-efficacy through design would encourage them to use SSK.

Older adults may also be more susceptible to situational deterrents such as perceived crowding, implied time pressure, and social presence of co-actors or non-interactive strangers when using SSK. The situational factors are known to invoke technology anxiety which lowers the intention to use SST (Gelbrich & Sattler, 2014;

Argo et al., 2005; Dabholkar & Bagozzi, 2002; Simon & Usunier, 2007; Wang et al., 2012) and negatively affect task performance especially when the task is complex and not well-learned (Zajonc, 1965). Time pressure, in particular, becomes intensified when the task is unfamiliar (Rastegary & Landy, 1993). As using SSK is rather a complex and unfamiliar task for older adults, the situational effects would be substantial for them.

## Self-service kiosks design for the aged population

Multiple previous studies suggested ways to accommodate older users' cognitive and physical characteristics in designing SSK. The suggested solutions to the cognitive aspects of difficulties include providing clear adjustable outputs with the use of appropriately sized text and high contrast colors (Charness et al., 2010; Hawthorn, 2000), and managing system complexity by controlling the number of levels of the menu to a minimum (Sharma et al., 2016; Czaja, 1996), or simplifying functionality based on the task demands and needs for services of older adults (Chan et al., 2009). Also, some studies discussed the importance of using appropriate feedback in order to minimize the demands on working memory (Caparni et al., 2012; Lee & Zhai, 2009). Caparni et al. (2012) noted the need for proper feedback to inform users of where they are in the system and where they have been. Modalities of feedback have also been an important topic. Some studies attempted to determine the modality or modality combination that increases ease of use (e.g., Tsai & Lee, 2009; Hwangbo et al., 2012). Furthermore, novel interactions have been proposed to aid older users who lack interaction abilities due to the absence of proper mental models. In the work of Mäkinen et al. (2002), a newly developed SSK incorporated a camera that recognized a user's face and facial expressions to help those who do not know what to do in real-time. Another study by Niculescu et al. (2016) developed a multimodal SSK which included natural language processing. The proposed SSK enabled users to use speech input with their smartphones in addition to conventional touch operations, thus encouraging them to verbally ask for help whenever troubled. In fact, multi-modality was found to increase the functionality and flexibility of SSK by taking advantage of diverse types of input and output during interactions (Günay & Erbuğ, 2015).

A number of studies have examined the design of SSK user interfaces for accommodating older users' physical characteristics. Multiple studies investigated the effects of target size, button spacing, and inclination angle on user performance and preference during the use of touchscreen interfaces (Colle & Hiszem, 2004; Gao & Sun, 2015; Sesto et al., 2012; Kim et al., 2014; Jin et al., 2007; Murata & Iwase, 2005; Schedlbauer, 2007). An intelligent user interface in which the SSK height and the sizes and locations of texts and buttons adapt to the user's physical characteristics has been proposed (Hagen & Sandnes, 2010) - from the user-centered design perspective, adjustable products that accommodate individual users' unique anthropometric characteristics benefit user experience and health outcomes (Dianat et al., 2018). In addition, some studies have explored the effects of posture (sitting or standing) on touchscreen task performance (Schedlbauer et al., 2006; Chourasia et al., 2013). Chourasia et al. (2013) reported that in general, the frequency of misses and task completion time were higher for standing than sitting, thus suggesting that postural conditions be considered when discussing the accessibility and usability of SSK.

Finally, multiple studies have developed user interface design guidelines for SSK that address different dimensions of difficulties of the older population (e.g., Maguire, 1999; Phiriyapokanon, 2011; Motti et al., 2013). Most guidelines discussed the importance of system simplicity, proper input and output, and feedback design, and, emphasized the need for customization. The existing guidelines, however, generally did not include recommendations on physical design.

# METHOD

## Participants

Two groups of participants, the younger and older groups, participated in this research study. The younger group (20~35 years old) consisted of twelve males and ten females; the older group (55~70 years old), fourteen males and thirteen females. Table 1 presents a summary of the participants' demographic data. At the time of the study, all participants were free of obvious musculoskeletal disorders and had a normal or corrected-to-normal vision in both eyes.

The research complied with the American Psychological Association Code of Ethics and was approved by the Institutional Review Board at Seoul National University. Informed consent was obtained from each participant.

Table 1. Summary of age, technology familiarity, and frequency of SSK usage (last 3 months) for the two participant age groups.

| Dimension | | Younger | Older |
|---|---|---|---|
| Age | (years) | 23.7 ± 2.1 | 61.9 ± 4.4 |
| Technology familiarity | (10-point scale) | 8.0 ± 0.7 | 2.4 ± 0.8 |
| Number of SSK uses in the last 3 months | | 8.7 ± 1.7 | 1.3 ± 1.2 |

*Note: Technology familiarity was assessed utilizing the 10-point frequency scale from Rosen et al. (2013).

## Experimental setup

An experimental setup was created to closely approximate the real usage situation in which people stand in a queue in front of SSK and take turns to order food items using the machine. Two identical SSK were installed in a row in a laboratory. Each consisted of a 23-in touch screen monitor, which is widely used in many restaurants in South Korea, and a height-adjustable monitor stand. The laboratory was spacious enough to allow much space ahead of the SSK so that participants could line up in front of them. In addition, three physical design features, side partitions, a back partition, and a chair, were prepared – they were used to construct different physical design alternatives. Three design variables, Side Partition, Back Partition, and Chair, were used in combination to represent different physical design possibilities of SSK. Each of these physical design variables had two levels: Yes (use of the feature) and No (no use).

## Physical design alternatives of self-service kiosks

The current study comparatively evaluated a total of eight physical design alternatives of SSK, which were the combinations of the levels of the three physical design variables: 8 = two levels of Side Partition (Yes/No) × two levels of Back Partition (Yes/No) × two levels of Chair (Yes/No). The participants performed the experimental task (a food ordering task) in each of the eight physical design alternatives. With Side Partition, Back Partition, and Chair represented as alphabet letters S, B, and C, the alternatives were indicated as S1B1C1, S0B0C1, etc., where the number right to each alphabet symbol indicated whether or not the physical design feature was used (1: Yes, 0: No). For instance, S1B1C1 indicates the design where there were side and back partitions and chairs (Figure 1a), while S0B0C1 represents the design where there were only chairs (Figure 1b).

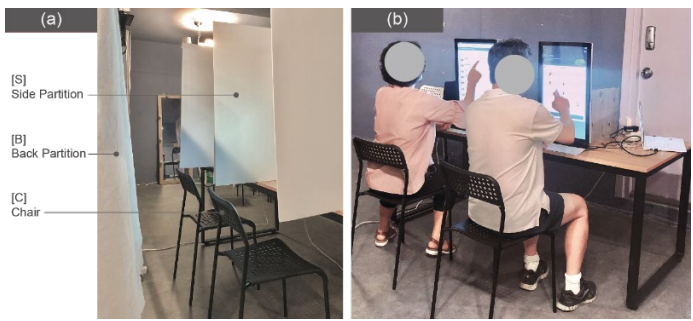

Figure 1. Experimental setup. (a) Three physical design features. (b) Participants performing the experimental task in S0B0C1.

## Task

The experimental task was a food ordering task. Each participant ordered a food item on the menu using the SSK in each of the eight physical design alternatives Each participant performed a single trial for each physical design alternative and thus a total of eight experimental trials. In order to motivate the participants to quickly complete the food ordering task as they usually do in fast-food restaurants, they were informed that the entire round of the experiment would finish as soon as everyone in the group completed the task. In each trial, each participant was given an instruction sheet on which a food item for the trial was written. The food items of the eight trials were all different for each participant.

The food ordering task consisted of five sub-tasks: selecting the burger, meal size, toppings, side dish, and beverage. For each sub-task, task accuracy was measured in terms of correct (Success) or incorrect (Failure) selection.

The BURGER KING®KOREA software application available from Google Play was used for the task. The Bandicam screen recorder and the open-source multimedia manipulation tool, FFmpeg, were used along with the Python script for data collection.

## Procedure

The participants were assigned to one of eight groups and participated in the experiment following the schedule of the group. The eight groups had similar age and gender distributions. The order of the eight experiment trials in each group was designed according to the Latin square design to ensure a counterbalanced experiment. Seven of the eight groups consisted of six participants; the other group, seven.

All participants participated in the research study for two days. On the first day, they were surveyed on their past SSK use experience and were given explanations about the experimental task. The task was fully explained to and practiced by the participants before the beginning of the first trial. In each trial, the participants were instructed to line up in front of the SSK in a single queue and use either of the two which was available on their turns. At the beginning of each trial, the instructor handed over the instruction sheet on which the burger, meal size, toppings, side dish, and beverage to be selected were written. The participants returned to the end of the line when they completed the food ordering task. Immediately after the completion of the task by all participants in the group, the participants completed the NASA-TLX questionnaire and took a rest for 15-20 minutes. The process, which required about 30 minutes, was repeated four times a day, a total of eight times in two days. When the participants completed all trials, they were thanked and debriefed (Figure 2). In order to control order effects and sequence effects, the positions of the participants in each queue were also determined according to the Latin square design.

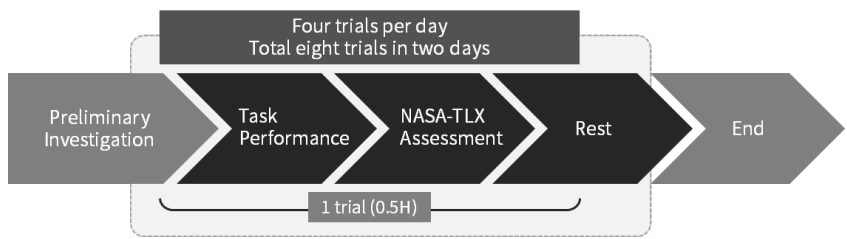

Figure 2. Experimental procedure.

## Independent and dependent variables

In this study, three independent variables (Side Partition, Back Partition and Chair) and a stratification variable (Age) were considered. Three independent variables (physical design variables) were within-subjects factors.

Each participant's task performance was evaluated in terms of task completion time and accuracy. Three task performance measures, that is, task completion time, average reaction time, and the number of sub-tasks completed successfully were employed as dependent variables (Table 2). In addition to the performance measures, the workload of each participant was evaluated utilizing the National Aeronautics and Space Administration Task Load Index (NASA-TLX) (Hart & Staveland, 1988). Each participant rated the perceived workload for each of the six dimensions of workload (mental demand, physical demand, time pressure, effort, performance, and frustration). The weighted average workload reflecting each participant's view on the relative contributions of the six dimensions was also calculated. Figure 3 presents a graphical summary of the independent/stratification and dependent variables.

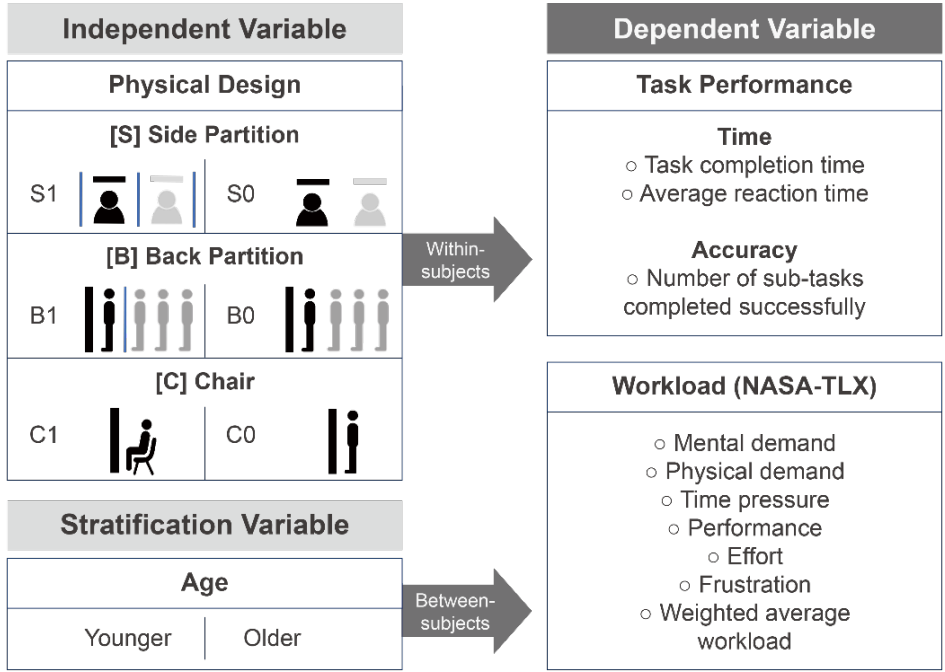

Figure 3. Summary of independent/stratification and dependent variables.

Table 2. Task performance measures.

| Task Performance Measure | | Definition/Quantification |
|---|---|---|
| Time | Task completion time | The total amount of time between the onset of the trial and completion of the last operation (s) |
| | Average reaction time | Task completion time/(Total number of touch operations -1) (s) |
| Accuracy | Number of sub-tasks completed successfully | Full marks: 5<br>The score was computed as the number of correctly performed sub-tasks. The five sub-tasks consisted of selecting: (a) burger, (b) meal size (regular or large), (c) toppings, (d) side dish and (e) beverage. |

## Data Analyses

The task performance data were examined for outliers prior to statistical analyses. Among the participants with task performance measures greater than 3 SD above or below the mean (Plumert & Schwebel, 1997), two participants of the older group were removed since their time and accuracy measures were both 3 SD below the mean. As their exceptionally low task accuracy with short task completion time implies careless participation, all of their data points were removed. Following outlier removal, for each dependent variable except the number of sub-tasks completed successfully, the study employed a general linear model of repeated measures to test the statistical significance of the independent/stratification variables and their interactions; a generalized linear model was employed for the number of sub-tasks completed successfully. The model included one between-subjects factor of 2 levels (Age: younger, older) and three within-subjects factors (Side Partition, Back Partition, Chair) of 2 levels (Yes, No). In cases where a higher-order interaction was statistically significant, post hoc tests with Bonferroni corrections were conducted. In addition, for each age group, the eight physical design alternatives were compared with one another in terms of the mean values of the workload and task performance measures. Also, a Pearson correlation analysis was conducted for each pair of a workload and a task performance measure to test a possible linear relationship between them. All statistical analyses were conducted using SPSS 25 (IBM Corp., Armonk, USA), and an alpha level of 0.05 was utilized.

# RESULTS

## GLM analyses

**Task performance**

Task performance was evaluated in terms of task completion time and accuracy. Age differences were significant for all task performance variables. The older group's mean task completion time and mean average reaction time were significantly longer than those of the younger group. The mean number of sub-tasks completed successfully was also significantly smaller for the older group than the younger group (Table 3, Figures 4a~4c). Regarding *task completion time*, aside from the age difference, the Age × Chair interaction effect was significant ($F(1,45)=4.239$, $p<0.05$). In particular, the mean task completion time was significantly longer for C0 than C1 in the older group ($p<0.05$; Table 3, Figure 5a). As to *average reaction time*, the main effect of Chair ($F(1,45)=7.732$, $p<0.01$) and a two-way interaction of Age × Chair were significant ($F(1,45)=4.328$, $p<0.05$). The mean average reaction time was significantly longer for C1 than C0 in the older group ($p<0.001$; Table 3, Figure 5b). Concerning the *number of sub-tasks completed successfully*, no significant effects other than the age main effect were found.

**Workload**

The results of the statistical analyses for each dimension of workload (mental demand, physical demand, time pressure, performance, effort, and frustration) and for the weighted average for the six dimensions are presented in Table 3. Age differences were significant in all dimensions - the older group's mean ratings were significantly higher than those of the younger group (Table 3, Figure 4d).

In addition to the age differences, regarding *mental demand*, a two-way interaction of Side Partition × Back Partition was significant ($F(1,45)=5.089$, $p<0.05$), with post hoc analyses indicating that mental demand was higher in S0B0 than S0B1 ($p<0.05$; Table 3, Figure 6a). For *physical demand*, a two-way interaction of Side Partition × Chair was significant ($F(1,45)=5.582$, $p<0.05$). In particular, the mean physical demand rating was significantly greater for S0C0 than for S0C1 ($p<0.01$; Table 3, Figure 6b). As to *time pressure*, the mean rating was significantly greater for B0 than B1 ($F(1,45)=4.511$, $p<0.05$; Table 3, Figure 6c) and also for C0 than C1 ($F(1,45)=5.495$, $p<0.05$; Table 3, Figure 6d). Concerning *performance*, the mean rating was significantly greater for S0 than S1 ($F(1,45)=6.179$, $p<0.05$; Table 3, Figure 6e). In regard to *effort*, a two-way interaction of Side Partition × Back Partition ($F(1,45)=6.612$, $p<0.05$) and a three-way interaction of Age × Side Partition × Back Partition ($F(1,45)=4.776$, $p<0.05$) were significant. The significant differences were found between S0B0 and S1B0, between S1B1 and S1B0, and between S0B0 and S0B1 in the older group ($p<0.05$; Table 3, Figure 6f). In other words, the antagonist interaction was found between Side Partition and Back partition in the older group. In the case of the younger group, no significant results were found. As for *frustration*, the main effect of Chair ($F(1,45)=6.864$, $p<0.05$) and a two-way interaction of Chair × Age were significant (F(1,45)=4.643, p<0.05). In particular, the mean frustration rating was significantly greater for C0 than C1 in the older group (p<0.001; Table 3, Figure 6g). In addition, a two-way interaction of Side Partition × Chair was significant (F(1,45)=4.453, p<0.05), the mean rating was significantly greater for S0C0 than S0C1 (p<0.001; Table 3, Figure 6h). With regard to the *weighted average workload*, the mean score was significantly greater for C0 than C1 (F(1,45)=6.980, p<0.05; Table 3, Figure 6i).

Table 3. Summary of the significant main effects and interactions (*: p<0.05, **: p<0.01, ***: p<0.001).

| | Dependent variable | Main effect | | | | Interaction effect | | | |
|---|---|---|---|---|---|---|---|---|---|
| | | Age | Side | Back | Chair | Age× Chair | Side× Back | Side× Chair | Age× Side× Back |
| Task performance | Task completion time | *** | | | | * | | | |
| | Average reaction time | *** | | | * | * | | | |
| | Number of sub-tasks completed successfully | *** | | | | | | | |
| Workload | Mental demand | *** | | | | | * | | |
| | Physical demand | *** | | | | | | * | |
| | Time pressure | ** | | * | * | | | | |
| | Performance | *** | * | | | | | | |
| | Effort | ** | | | | | * | | * |
| | Frustration | *** | | | * | * | | * | |
| | Weighted average workload | *** | | | * | | | | |

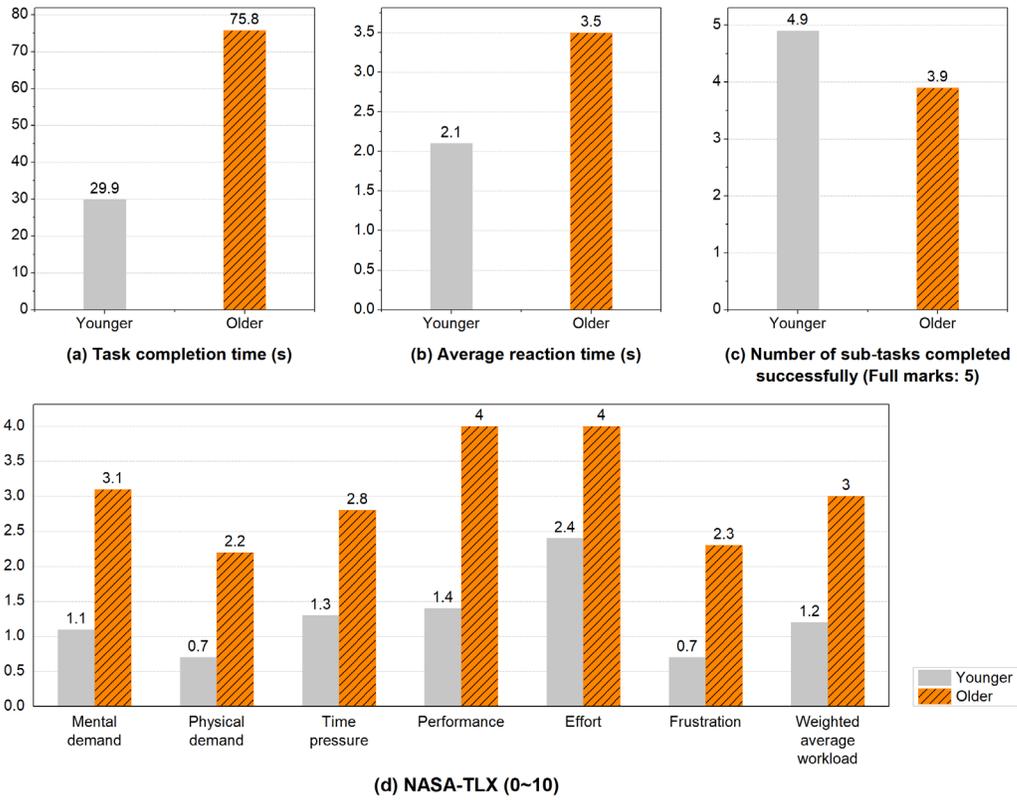

Figure 4. Descriptions of age differences.

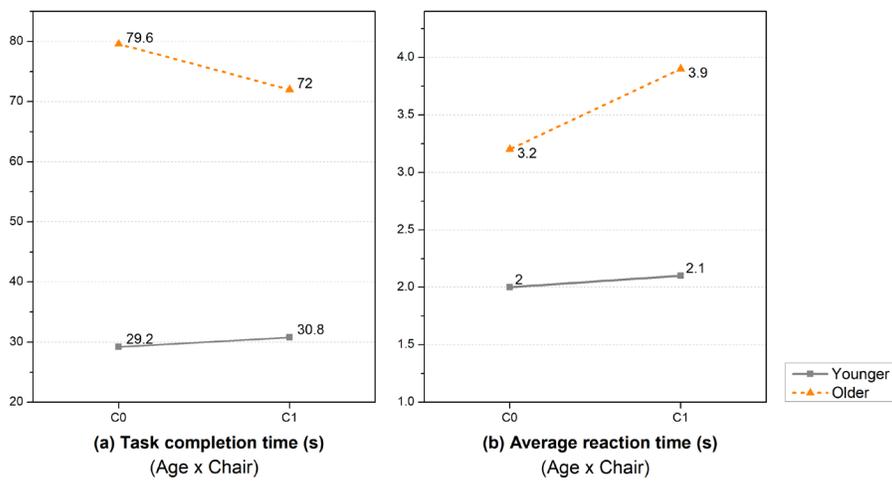

Figure 5. Descriptions of significant results for task performance variables.

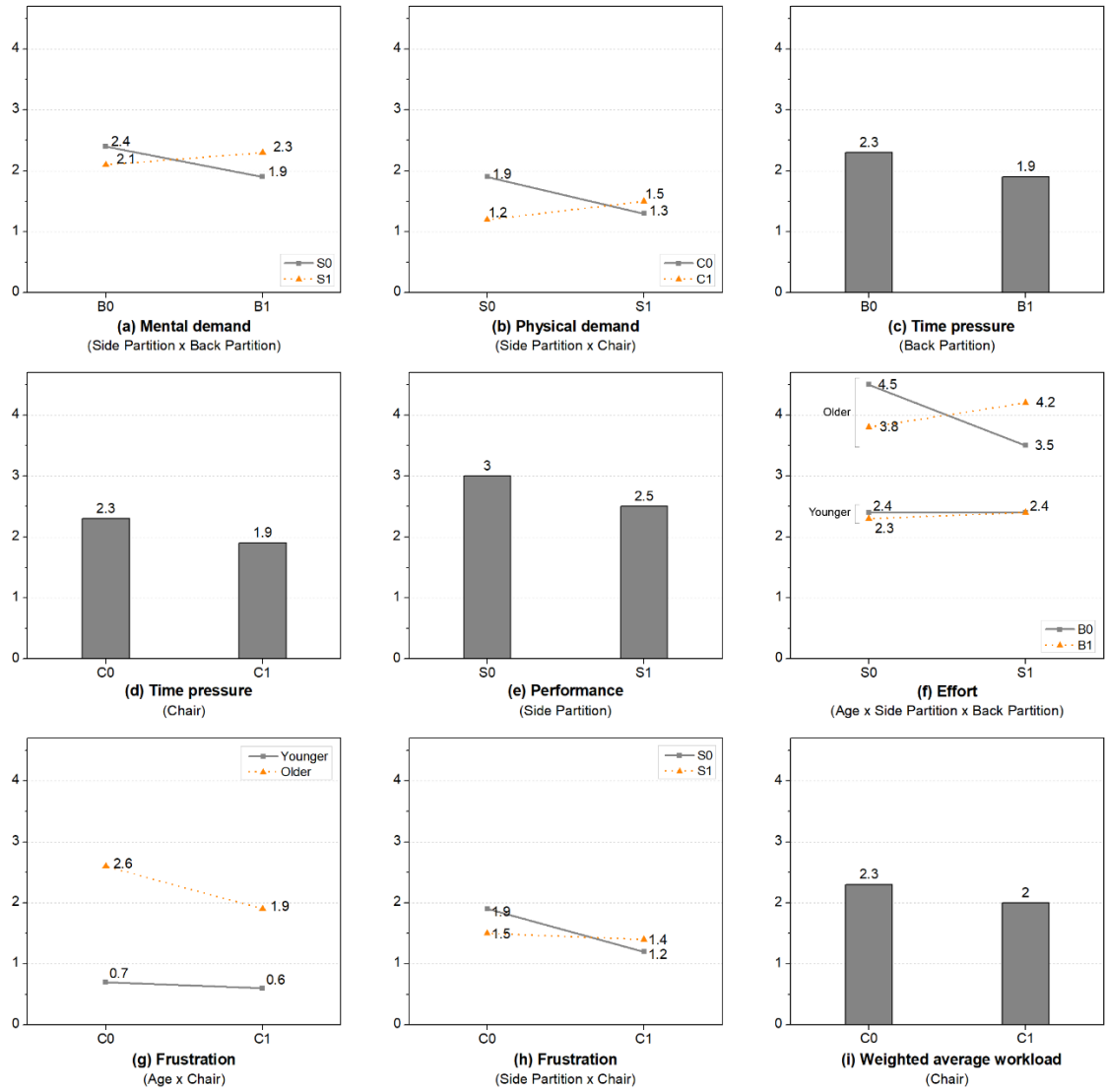

Figure 6. Descriptions of significant results for workload variables.

**Comparison of design alternatives**

For each age group, the eight physical design alternatives were compared with one another in three evaluation criteria – the mean weighted average workload score, mean task completion time, and mean number of tasks completed successfully. Figure 7 presents a graphical summary of the comparisons; the x-axis and y-axis respectively indicate the mean task completion time and the mean number of sub-tasks completed successfully while the mean weighted average workload of each design alternative is written in the vicinity of the corresponding symbol.

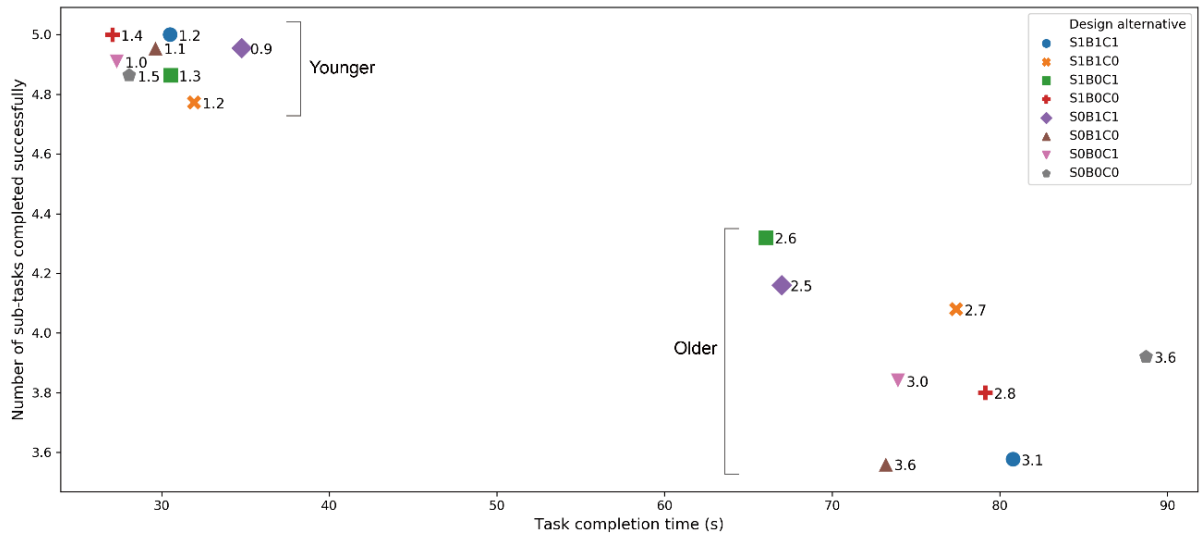

Figure 7. Comparisons of the eight physical design alternatives.

## Correlation analyses

Pearson correlation coefficients between the workload and the task performance variables were computed for each age group (Table 4). Table 4 indicates that overall, linear correlations between the workload and the task performance variables were more pronounced for the older group (Table 4b) than for the younger group (Table 4a).

Table 4. Correlations between workload and task performance variables (*: $p<0.05$, **: $p<0.01$; Correlation coefficients larger than 0.3 were highlighted).

(a) Correlations of the younger group.

| Dependent variable | | Workload | | | | | | |
|---|---|---|---|---|---|---|---|---|
| | | Mental demand | Physical demand | Time pressure | Performance | Effort | Frustration | Weighted average workload |
| Task performance | Task completion time | 0.29** | | 0.29** | 0.31** | 0.34** | 0.38** | 0.39** |
| | Average reaction time | | 0.18* | | | | | |
| | Number of sub-tasks completed successfully | | | | | -0.22** | | |

(b) Correlations of the older group.

| Dependent variable | | Workload | | | | | | |
|---|---|---|---|---|---|---|---|---|
| | | Mental demand | Physical demand | Time pressure | Performance | Effort | Frustration | Weighted average workload |
| Task performance | Task completion time | 0.42** | 0.34** | 0.44** | 0.25** | 0.44** | 0.43** | 0.48** |
| | Average reaction time | | | | | | | |
| | Number of sub-tasks completed successfully | -0.40** | -0.29** | -0.26** | -0.30** | -0.25** | -0.40** | -0.39** |

# DISCUSSION

The research study elucidated the effects of privacy partitions and chairs on task performance and workload during an SSK task for two age groups. The SSK task was a typical food ordering task. A total of eight design alternatives (8 = two levels of Side Partition × two levels of Back Partition × two levels of Chair) were considered.

## Age and self-service kiosks (SSK)

The two age groups differed significantly in all dependent variables. The mean task completion time and the average reaction time were longer, the number of sub-tasks completed successfully was smaller, and the mean workload scores of all dimensions were higher for the older group than for the younger (Figure 4). In other words, the older participants on average required more time to complete the task, and, encountered more difficulties when completing the sub-tasks.

The observed age impacts on the performance measures are thought to be due to the cognitive, physical, attitudinal and situational characteristics of older adults. Age-associated declines in visual information processing and motor control (Walsh & Prasse, 1980; Cerella et al., 1982; Iancu, I. & Iancu, B., 2017) explain the increases in the time-based performance measures. Similarly, age-related declines in divided attention and executive control (Craik & McDowd, 1987; Rabbit, 2019; Phillips & Sternthal, 1977) explain the decreased numbers of sub-tasks completed successfully. Lower confidence of older adults participants in their ability to complete the tasks and overcome situational deterrents (Gelbrich & Sattler, 2014; Wang et al., 2012) may also explain the lower performance outcomes observed. The motivation theory (Fagan et al., 2008) argues that both extrinsic and intrinsic motivations influence human performance or capabilities. Thus, factors, such as increased anxiety and lower confidence, may have negatively affected the older participants' task performance.

The study results on the workload measures also appear to reflect the cognitive, physical, attitudinal and situational characteristics of older adults. These characteristics can be understood as reductions in different mental and physical resources and in the ability to cope with different stressors. Workload is defined as the discrepancy between task demands and available resources (Hart & Staveland, 1988). Accordingly, reductions in resources would naturally lead to increases in workload in different dimensions, as observed in this study. The age impacts observed for all six dimensions of the NASA-TLX suggest that such reductions in resources occur across various dimensions. Relatedly, it is worth noting the results on the physical demand and time pressure measures - they seem to confirm the importance of providing physical supports and measures for alleviating situational stressors for older SSK users.

## Effects of SSK physical design by age

Regarding the task performance measures, the significant two-way interactions of Age × Chair on task completion time (Figure 5a) and average reaction time (Figure 5b) indicate that: 1) across the eight experimental trials, the younger group mostly performed the task well in 30 seconds irrespective of physical design; 2) in contrast, the older group's task completion time was more dependent on the use of chairs; 3) sitting enabled the older group to complete the task more quickly (Figure 5a) with longer reaction times (Figure 5b). In contrast, when standing, the older participants tended to perform individual touch operations more quickly, seemingly in an attempt to reduce task completion time, although this indeed resulted in increasing it.

Concerning the workload measures, several significant main and interaction effects of the physical design variables were found for the mental demand (Figure 6a), physical demand (Figure 6b), time pressure (Figures 6c and 6d), performance (Figure 6e), frustration (Figure 6h), and weighted average workload (Figure 6i) measures. Meanwhile, two interaction effects involving Age, that is, Age × Side Partition × Back Partition and Age × Chair, were found for effort (Figure 6f) and frustration (Figure 6g), respectively. Each of these interaction effects indicates that the corresponding physical design variable or combination of physical design variables affected the workload measure only in the older group. Overall, the results concerning the workload measures imply that 1) making use of chairs not only benefitted reducing task completion time for the older group (Figure 5a) but also decreased the weighted average workload for both age groups and that 2) the use of side partitions and a back partition were respectively effective in terms of improving performance and reducing time pressure, but concerning mental demand and effort, using only either one of the two features was beneficial as compared to using both.

As for *mental demand*, the Side Partition × Back Partition interaction effect suggests that in the absence of side partitions, the condition without a back partition resulted in a significantly higher mean mental demand score compared to the condition with it (Figure 6a). Following the definition of social presence, that is, 'the degree of salience of the other' (Short et al., 1976), the combinations of Side Partition and Back Partition (S1B1, S1B0, S0B1, and S0B0) could be classified into three categories representing different degrees of social exposure: S1B1 (no exposure), S1B0 and S0B1 (medium exposure), and S0B0 (full exposure). The antagonistic two-way interaction seems to imply that the medium degree of social exposure is beneficial in terms of mental demand compared to no or full exposure. It may be that the trade-off between anxiety from social pressure and a feeling of isolation renders the medium exposure optimal.

Regarding *physical demand*, the Side Partition × Chair interaction effect indicates that in the presence of side partitions, the sitting and standing conditions did not differ in terms of the mean physical demand score, whereas, in the absence of side partitions, the standing condition resulted in a significantly higher mean physical demand score than the sitting condition (Figure 6b). It is thought that the side partitions served as visual references that facilitated postural control during standing and that their absence, therefore, increased the difficulty and physical workload associated with standing.

Related to this, Teasdale et al. (1993) reported that a decrease in the amount of available sensory information increased the difficulty of postural control during standing, which increased postural sway. Also, Simeonov & Hsiao (2001) and Simeonov et al. (2009) demonstrated the effectiveness of visual references for balance improvement during standing.

With regard to *time pressure*, the installation of a back partition and chairs both decreased the perceived level of time pressure (Figures 6c and 6d). Consciously knowing that other participants were waiting for their turns likely put time pressure on the participants as they performed the task; the installation of a back partition is thought to have reduced the awareness of others waiting behind. The main effect of Chair on time pressure indicates that the participants felt less rushed when they were seated. The concept of affordance, that is, perceivable action possibilities (Norman, 1988) may explain this result. The presence of the chair itself may have led the participants to perceive that they could complete the task in a more relaxed state taking ample time.

Regarding *performance*, the main effect of Side Partition (Figure 6e) indicates that the sheltering of co-actors was helpful for the participants, as they felt more satisfied with their task performance. As argued by Seta (1982), when there are co-actors, and the tasks are identical, individuals tend to attribute any differences in performance levels to differences in abilities. Such a tendency could increase the level of anxiety and decrease the confidence level of a performer. It may be that the use of side partitions helped enhance the confidence of the participants by making any self-comparison impossible.

As for *effort*, an antagonistic interaction effect between the two types of partitions similar to that observed for mental demand was found only in the older group (Figure 6f). The older participants felt that they put less effort into the task when they were exposed to the medium degree of social exposure as compared to none or full. This result may also suggest that an appropriate level of social presence minimizes perceived task difficulty by maximizing social facilitation (Zajonc, 1965).

Concerning *frustration*, the significant Age × Chair interaction effect indicates that the use of a chair decreased frustration especially in the older group (Figure 6g). Compared with younger adults, older adults are known to allocate more attention to postural control during standing (Lajoie et al., 1993; Lajoie et al., 1996; Remaud et al., 2012; Lacour et al., 2008). It is thought that the older participants felt significantly less frustration when seated than standing because sitting required less attention than standing. The result is also in line with the age differences in physical demand (Figure 4d), thus emphasizing the need to minimize older adults' physical workload when they perform a cognitive task.

The greater mean *frustration* for S0C0 than for S0C1 shown in Figure 6h also illustrates the benefits of using a chair, especially in the absence of side partitions. This result may be interpreted in relation to the Side Partition × Chair interaction effect on physical demand illustrated in Figure 6b – the side partitions are thought to have served as visual references for postural control; thus, a chair was likely more beneficial in reducing physical demand and thus alleviating frustration in the absence of the visual partitions.

Finally, regarding the *weighted average workload*, the main effect of Chair indicates that the sitting condition lowered it for both groups (Figure 6i). The results shown in Figure 6 suggest that the use of a chair reduced time pressure for both participant groups (Figure 6d), and, for the older group, it also lowered the level of frustration. These chair effects in combination are thought to have contributed to the observed chair effect on weighted average workload (Figure 6i).

The benefits of a chair for older adults observed in the study are in line with past research studies, which emphasized that postural control is not a simple interplay between static reflexes but is rather a complex skill supervised by high-level cognitive processes (Demanze & Michel, 2017; Remaud et al., 2012; Lacour et al., 2008). With normal aging, based on the 'posture first' principle (Andersson et al., 2002), cognitive task performance deteriorates during dual-task situations involving a postural and a cognitive task. This occurs because balance is prioritized, and the maintenance of balance requires more central processing with a greater amount of attentional resources allocated to the postural task (Ruffieux et al., 2015; Lajoie et al., 1996). Consequently, for older adults, minimizing the postural control demand through the use of a chair would decrease the divided attentional costs, and, thus, would contribute to improving task performance and lowering workload across different dual-task situations.

## SSK physical design recommendations for the aged population

Overall, the study proposes two SSK design recommendations: 1) place chairs so that older adults can operate SSK while sitting, and 2) make use of either side partitions or a back partition to create a medium degree of social presence. The first design recommendation is to help older users perform SSK tasks more quickly and with less frustration. Also, it would well benefit users of all ages by reducing the overall workload. The second design recommendation is to reduce workloads during the task in the mental demand and effort dimensions.

The comparisons of the eight physical design alternatives in terms of the average task completion time and the average accuracy and workload scores (Figure 7) also support the design recommendations. For the older group, S1B0C1 and S0B1C1 were the two most desirable alternatives when considering the three criteria in combination. S1B0C1 resulted in the shortest mean task completion time (66.1s), the largest mean number of sub-tasks completed successfully (4.3), and the second-lowest mean weighted average workload score (2.6). Another comparable alternative, S0B1C1, resulted in the second-shortest mean task completion time (67.0s), the second-largest mean number of sub-tasks completed successfully (4.2), and the lowest mean weighted average workload score (2.5). Compared to the control group where none of the physical design features were applied (S0B0C0), S1B0C1 and S0B1C1 reduced the mean task completion time by more than 20 seconds while increasing accuracy and reducing workload at the same time. In order to test these differences, additional one-way repeated measures ANOVA were conducted. Regarding the task completion time, the physical design alternative effect was significant in the older group, and the mean task completion times of S1B0C1 and S0B1C1 were significantly shorter than that of S0B0C0. The results overall recommend S1B0C1 or S0B1C1.

## Implications

To the best of our knowledge, this study is the first to empirically demonstrate the difficulties experienced by older users when using SSK with both objective and subjective data. The study results lend support to past research studies about the physical, attitudinal and situational aspects of problems experienced by older adults, and suggest that simple design changes that provide environmental support for declining abilities would serve as a powerful intervention to improve the performance of older adults by resolving the problems (Charness, 2008). If the effects of physical design were integrated with GUI improvements, it would greatly contribute to decreasing the gap between the younger and older groups.

Designing SSK to increase the sense of consideration felt by older users would also encourage them to become more engaged in the technological age (Günay & Erbuğ, 2015). In general, the ability to use technology improves the independence of older adults and their perception of the quality of life (Mynatt & Rogers, 2002). Therefore, promoting their use of SSK through design improvements would mean more than enhanced usability.

In addition, it should be noted that SSK design could provide benefits to the business side. Shorter ordering times and higher accuracy levels would not only lead to shorter cycle times and lower operational costs but would also attract more older adults customers. Once older adults gain some positive experience with SSK, their tendency to avoid using them would decrease, and they would more eagerly visit restaurants to reuse SSK (Zhao et al., 2008). Therefore, small physical design modifications could help boost many businesses' revenue levels.

## Limitations and future work

This study was a lab experiment; thus, it is possible the actual effects of the physical design variables in the real-world environments do not equal those reported in this study. We expect that in actual situations with more disturbances such as different visual signs and noises and where people are more willing to order as soon as possible to receive their food quickly, the effects of physical design features would likely be more significant. Observational studies in real-world settings are needed to confirm this prediction. Currently, such an observational study is under our investigation.

Also, since the task difficulty level was controlled throughout the experiment, investigating the impacts of task difficulty level would produce more complete knowledge. Future studies that explore more complicated SSK tasks, such as making payments and data entry, could help produce physical design recommendations for different task difficulty levels. The idea of adding physical design features could also be applied to diverse situations in which older adults must interact with different technological systems in a public space. In addition, the results overall imply that a medium degree of social exposure may help to control perceived difficulties of different tasks. Further investigations on the impacts of the degree of designed social exposure on the user experience of older adults are needed.

## CONCLUSION

Through a lab experiment, this study elucidated the impacts of physical design on workload and task performance during the use of self-service kiosks (SSK). This study also developed SSK physical design recommendations for the aged population. Notably, older adults conducted more effective touch operations and completed the SSK task more quickly when they were seated than standing. Making use of a chair also significantly reduced time pressure, frustration, and overall workload as evaluated using the NASA-TLX questionnaire. Moreover, creating the medium degree of social exposure by the use of either side or back partitions complemented the Chair effect by reducing the mean ratings of mental demand and effort. The results suggest that a chair should be positioned with the installation of either side partitions or a back partition to enhance the older users' experience during the use of SSK. The current study results also present a novel approach to bridging the digital divide. While growing technological developments and the widespread installation of SSK may threaten older adults who are less familiar with new technology, designing an older adults-friendly physical space around SSK would significantly help them use SSK more effectively and thus help them become more involved in the digital society.

## KEY POINTS

- The study results demonstrated that sitting (the use of a chair) while using self-service kiosks (SSK) would especially benefit older adults with regard to task performance and workload levels.
- Compared to standing, sitting resulted in nearly a 10% decrease in task completion time while also reducing workloads, including time pressure and frustration perceived by the older group. The observed benefits of a chair may be attributed to reduced postural control demands and divided attention costs

- The use of either side or back partitions complemented the chair effect; S1B0C1 and S0B1C1 resulted in about a 24% decrease in task completion time compared to S0B0C0 in the older group. They also yielded greater numbers of sub-tasks completed successfully and lower weighted average workload scores.
- The use of the proposed physical design recommendations would greatly help older adults use SSK more effectively.

## Biographies

Hyesun Chung is currently a graduate student researcher in Industrial Engineering at Seoul National University in Seoul, South Korea. She received her bachelor's degrees in science in industrial engineering, fine arts in design, and business administration from Seoul National University.

Woojin Park is currently a professor in Industrial Engineering at Seoul National University in Seoul, South Korea. He received his B.S. and M.S. degrees in Industrial Engineering from POSTECH and Ph.D. degree in industrial and operations engineering from the University of Michigan, Ann Arbor.